\documentclass[graybox]{svmult}
\usepackage{helvet}         
\usepackage{graphicx}        
\begin{document}

\title{A method to measure vacuum birefringence at FCC-ee}
\author{Ulrik I. Uggerh{\o}j and Tobias N. Wistisen}
\institute{ \at Department of Physics and Astronomy, Aarhus University, Denmark}
%
%
\maketitle

\abstract{
It is well-known that the 
Heisenberg-Euler-Schwinger effective Lagrangian predicts that
a vacuum with a strong static electromagnetic field turns birefringent. We propose a scheme that can be implemented at the planned FCC-ee, to measure the nonlinear effect of vacuum birefringence in electrodynamics
arising from QED corrections.
Our scheme employs a pulsed laser to create Compton backscattered
photons off a high energy electron beam, with the FCC-ee as a particularly interesting example. These photons will pass through
a strong static magnetic field, which changes the state of polarization of the radiation - an effect proportional
to the photon energy. This change will be measured by the use of an aligned
single-crystal, where a large difference in the pair production cross-sections can be achieved.
In the proposed
experimental setup the birefringence effect gives rise to
a difference in the number of pairs created in the analyzing crystal,
stemming from the fact that the initial laser light has a varying state of polarization, achieved
with a rotating quarter wave plate. Evidence for the vacuum birefringent effect will be seen as a distinct peak
in the Fourier transform spectrum of the pair-production rate signal.
This tell-tale signal can be significantly above background with only few hours of measurement, in particular at high energies.
}

\section{Introduction}
\label{sec:1}
In this paper we employ natural units $\hbar=c=1$, $\alpha=e^{2}$.
The Heisenberg-Euler-Schwinger effective Lagrangian describes the
electromagnetic fields while keeping only the QED corrections to lowest
order in the fine structure constant, but including any (even) number
of photons. The result was first derived by W. Heisenberg and H. Euler.
\cite{EWH}. We use the expression of J. Schwinger \cite{PhysRev.82.664}
which has an asymptotic expansion of the Lagrangian:

\begin{equation}
\mathcal{L}=\mathcal{F}+\frac{\alpha^{2}}{90\pi m^{4}}\left[4\mathcal{F}^{2}+7\mathcal{G}^{2}\right]+...,\label{eq:f=0000F8rsteordenlagrangian}
\end{equation}
If one considers the situation with two fields, a strong static ``background''
field, and a weak perturbing radiation field, one can from the above
Lagrangian derive the field equations for the radiation field. To leading order in the ``background'' fields this manifests itself,
as if the radiation were propagating in a dielectric medium, with the
permittivity and permeability depending on the strong background fields

\begin{eqnarray}
\epsilon_{ik} & = & \delta_{ik}+\frac{\alpha^{2}}{45\pi m^{4}}\left[2(\mathbf{E}^{2}-\mathbf{B}^{2})\delta_{ik}+7B_{i}B_{k}\right],\label{eq:permittivitet}
\end{eqnarray}

\begin{equation}
\mu_{ik}=\delta_{ik}+\frac{\alpha^{2}}{45\pi m^{4}}\left[2(\mathbf{B}^{2}-\mathbf{E}^{2})\delta_{ik}+7E_{i}E_{k}\right],\label{eq:permeabilitet}
\end{equation}
as seen in \cite{jackson91} or \cite{beresteckij_quantum_2008}.
The field quantities in these dielectric tensors are the ones from
the strong background field, which we here take as being purely magnetic in the laboratory frame. Solutions are readily obtained which
yield a difference in the refractive index, depending on whether the
polarization of the radiation is in the same direction as the magnetic
field, or perpendicular to it. The results are

\begin{equation}
n_{\bot}=1+\frac{7\alpha}{90\pi}\frac{\mathbf{B}^{2}}{B_{c}^{2}},\label{eq:northo}
\end{equation}

\begin{equation}
n_{\Vert}=1+\frac{2\alpha}{45\pi}\frac{\mathbf{B}^{2}}{B_{c}^{2}}.\label{eq:npara}
\end{equation}
where $B_{c}=\frac{m^{2}}{e}$ is the Schwinger critical field ($4.4\cdot10^{9}\textrm{T})$.
Formulas (\ref{eq:northo})
and (\ref{eq:npara}) are the low frequency limit of the general result \cite{beresteckij_quantum_2008},\cite{Shore2007219},
valid as long as
$\omega\ll m\frac{B_{c}}{B}$ , where $B$ is the strength of the
static magnetic background field. Our proposal operates far below this limit.
The difference in refractive indices induces a phase shift of the radiation
between the two polarization directions given by

\begin{equation}
\Delta=\omega L_{B}\frac{3\alpha}{90\pi}\frac{\mathbf{B}^{2}}{B_{c}^{2}},\label{eq:deltan}
\end{equation}
with $L_{B}$ being the length of the dipole magnet and $\omega$
the photon energy.

\section{Experimental setup}
\label{sec:2}

\begin{figure}[t]
\sidecaption[t]
\includegraphics[width=\textwidth]{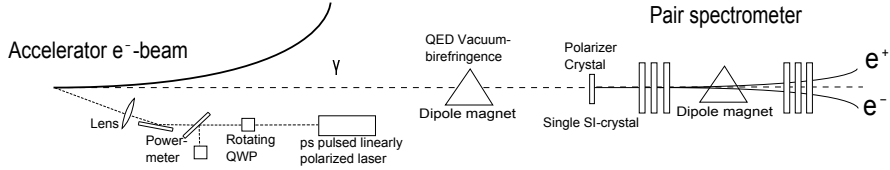}
\caption{Experimental setup\label{fig:Experimental-setup}. Linearly polarized
laser light passes through a rotating quarter wave-plate, the power
is measured from a small fraction of the light, and then the remaining light undergoes Compton backscattering. The backscattered
photons pass through the high field dipole magnet, and the resulting
radiation is analyzed using a properly oriented single Si-crystal followed by a pair spectrometer.}
\label{Experimental-setup}       
\end{figure}

The experimental setup proposed to measure this effect can be seen
in figure \ref{Experimental-setup}. In short, linearly polarized
laser light passes through a rotating quarter-wave plate, the power
is measured and then the light undergoes Compton backscattering from the intense and energetic electron beam. The electron beam must be of low emittance and short pulse-length where the laser pulse is matched with the duration of the $e^-$-beam. The backscattered
photons, now of very high energy, pass through a high field dipole magnet which for this example is taken as a standard LHC dipole, and the resulting
radiation is analyzed using a single Si-crystal and a pair spectrometer. The frequency of rotation of the initial quarter-wave plate then gives rise to a distinct peak in the Fourier spectrum of the number of pairs produced -- yielding a clear signal of the effect sought.

\subsection{Electron beam parameters}
\label{subsec:4}
For the electron beam, we choose the parameters of the FCC-ee as given in \cite{2014arXiv1411.2819K}. The FCC-ee is a circular machine of circumference 80-100 km which is presently under study, mainly at CERN, and it has several stages named after its main production potential, e.g.\ the \texttt{tt}-stage for top-production. The energies of the stages are $45.5$ GeV (\texttt{Z}-stage), $80$ GeV (\texttt{W}-stage), $120$ GeV (\texttt{H}-stage) and $175$ GeV (\texttt{tt}-stage).

\subsection{Modification of the polarization state}
\label{subsec:1}
To calculate the resulting pair creation
rate at the tracking detectors we use the formalism of M\"{u}ller calculus
\cite{Handbook} which facilitates the calculation of the polarization
and intensity of radiation through an optical system. The radiation
is described by a Stokes vector and the optical elements with a M\"{u}ller
matrix, for details see \cite {PhysRevD.88.053009}.
The final
Stokes vector is determined by operating on the initial Stokes vector
by the M\"{u}ller matrix of each component of the setup. In our setup,
we will have 4 M\"{u}ller matrices. One for a rotating quarter-wave-plate, one for the Compton backscattering process, one
for the QED process and one for the crystal polarizer. Thus, we obtain the intensity in each polarization component.

\subsection{Compton backscattering}
\label{subsec:2}

We employ the differential cross section for Compton
scattering for the initial polarization state in question, given in \cite{Ginzburg19845}, which is identical to -- but expressed differently from -- the Klein-Nishina cross section. Since the Klein-Nishina cross section applies to an electron at rest, two Lorentz-transformations are required to get the backscattering cross section and the emerging photon energies. 

\subsection{Pair production}
\label{subsec:3}
For the experimental analysis of the polarisation state, we utilize the state-dependent pair production rate from each of
the two orthogonal directions in the crystal. The transmittances are given as $\mathcal{T}_{\perp}=e^{-\sigma_{\bot}(y)\cdot L}$
and $\mathcal{T}_{\parallel}=e^{-\sigma_{\Vert}(y)\cdot L}$ with
$L$ being the traversed distance and $\sigma_{\bot}(y)$, $\sigma_{\Vert}(y)$
being the total number of pairs created per distance (the inverse of the mean free path), which depend
on the photon energy parametrized by $y=\omega_f/E$. To calculate the pair
production cross-sections we use the theory of coherent pair production, see e.g.\ \cite{TerMik}. The
differential pair production cross section depends on the asymmetry
between the energies of the two particles: $z=\frac{\varepsilon_{-}}{yE}$,
with $\varepsilon_{-}$ being the energy of the pair-produced electron.

The maximum photon energy $\omega_m$ of the Compton backscattered photons in the interaction of an electron beam of energy $E$ and a laser with photons of energy $\omega_0$ is given as
\begin{equation}
\omega_m=E\frac{x}{x+1}\qquad\mathrm{with}\qquad x=\frac{4E\omega_0}{m^2}
\end{equation}
which for $E=175$ GeV and $\lambda_0=1064$ nm, i.e.\ $\omega_0=1.165$ eV, yields  $x=3.12$ and thus $\omega_m=132.5$ GeV. The threshold for creation of pairs is $\omega_0\omega_m=m^2c^4$ which corresponds to $x^2/4(x+1)=1$ or $x=2(1+\sqrt{2})\simeq4.83$, so our scheme is still below the photon-photon pair-production threshold.

\begin{figure}[t]
\sidecaption[t]
\includegraphics[width=\textwidth]{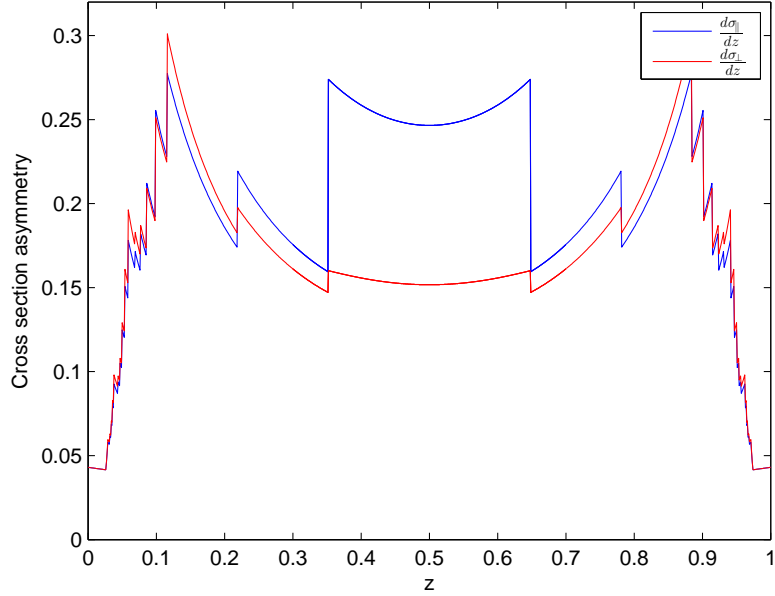}
\caption{Differential pair production inverse mean free path for Si at 132 GeV
photon energy. $\theta=1.4$mrad, $\alpha=0.16$, where $\theta$
is the angle between the momentum of the incoming particle $\mathbf{p}_{1}$
and the $[110]$ axis, and $\alpha$ is the angle between the plane
containing $\mathbf{p}_{1}$ and the $[110]$ axis with the plane
containing the axes $[001]$ and $[1\bar{1}0]$, see \cite{TerMik}}
\label{Differential-pair-production}       
\end{figure}

A plot of $\frac{d\sigma_{\bot}}{dz}$ and $\frac{d\sigma_{\Vert}}{dz}$,
for $132$ GeV incoming photons, can be seen in figure \ref{Differential-pair-production}.
The angles were chosen to achieve a significant asymmetry $\frac{d\sigma_{\Vert}-d\sigma_{\bot}}{d\sigma_{\Vert}+d\sigma_{\bot}}$
over the interval $0.3<z<0.7$ at a photon energy of $132$ GeV.

Setting the
degree of longitudinal polarization of the electron beam to zero $\lambda=0$
and carrying out the entire calculation yields
\begin{eqnarray}
\frac{d^{3}N_{\textrm{pairs}}}{dtdzdy}&=&\mathcal{L}_{e\gamma}\frac{d\sigma_{c}^{np}}{dy}\frac{I_{i}}{2\omega_{i}}[\left(1+\frac{2r^{2}}{f_{1}(y)}\cos^{2}\theta-\frac{f_{3}(y)}{f_{1}(y)}\Delta\sin\theta\right)\frac{1-q(y)}{\sigma_{\perp}(y)}\frac{d\sigma_{\perp}(y)}{dz}
\nonumber \\ 
&+&\left(1-\frac{2r^{2}}{f_{1}(y)}\cos^{2}\theta+\frac{f_{3}(y)}{f_{1}(y)}\Delta\sin\theta\right)\frac{1-r(y)}{\sigma_{\parallel}(y)}\frac{d\sigma_{\parallel}(y)}{dz}]
\label{eq:diffpaircalc}
\end{eqnarray}
where $y=\frac{\omega_{f}}{E}$ and $E$ is the total electron energy
and 
\begin{equation}
f_{1}(y)=\frac{1}{1-y}+1-y-4r(1-r),\label{eq:f1}
\end{equation}
\begin{equation}
f_{2}(y)=2\lambda rx[1+(1-y)(2r-1)^{2}],\label{eq:f2}
\end{equation}
\begin{equation}
f_{3}(y)=(1-2r)(\frac{1}{1-y}+1-y),\label{eq:f3}
\end{equation}
$\omega_{f}$ and $\omega_{i}$ are the photon energies after and
before the scattering process, respectively and $\theta$
is the angle between the momentum of the incoming particle $\mathbf{p}_{1}$
and the $[110]$ axis of the analyzer crystal.

The luminosity is given by
\begin{equation}
\mathcal{L}_{e\gamma}=2N_{e}\int\int\rho_{\gamma}(\vec{x},t)\rho_{e}(\vec{x},t)d^{3}\vec{x}dt,\label{eq:lumsortof}
\end{equation}
where $\rho_{e}(\vec{x},t)$ and $\rho_{\gamma}(\vec{x},t)$ are the
unity normalized density profiles of the electron bunch and laser
pulse, $N_{e}$ is the number of electrons in the bunch, $r=\frac{y}{x(1-y)}$
with $x=\frac{4E\omega_{i}}{m^2}$ as above, $\lambda$ is the degree of
longitudinal polarization of the electron beam while 
\begin{equation}
\frac{d\sigma_{c}^{np}}{dy}=\frac{2\pi\alpha^{2}}{xm^{2}}f_{1}(y),\label{eq:dsnp}
\end{equation}
\begin{equation}
\frac{d\sigma_{1}}{dy}=\frac{2\pi\alpha^{2}}{xm^{2}}rx(1-2r)(2-y),\label{eq:ds1}
\end{equation}
and $\sigma_{\parallel}$ and $\sigma_{\perp}$ are the Compton scattering cross sections, as seen in \cite{Ginzburg19845}.

\section{Results}
\label{sec:3}
If we now consider $\theta=\omega_{0}t$ and integrate
over the whole energy interval $0<y<y_{m}$ and integrate over a suitably
chosen interval for $z$ we get a pair-production rate. The Fourier
transform of this rate has components at frequencies $\omega=0$,
$\omega=\omega_{0}$ and $\omega=2\omega_{0}$. The component at $\omega=\omega_{0}$, selectable by tuning the frequency of the quarter-wave plate,
is the one of interest. It is only present when the magnet is turned
on, and thus signifies the effect of vacuum birefringence. The component
at the double frequency is due to the fact that the polarization state
of the Compton backscattered radiation depends on the initial polarization,
and the polarizer crystal turns this into a difference in pair production
rate.

\begin{figure}[t]
\sidecaption[t]
\includegraphics[width=\textwidth]{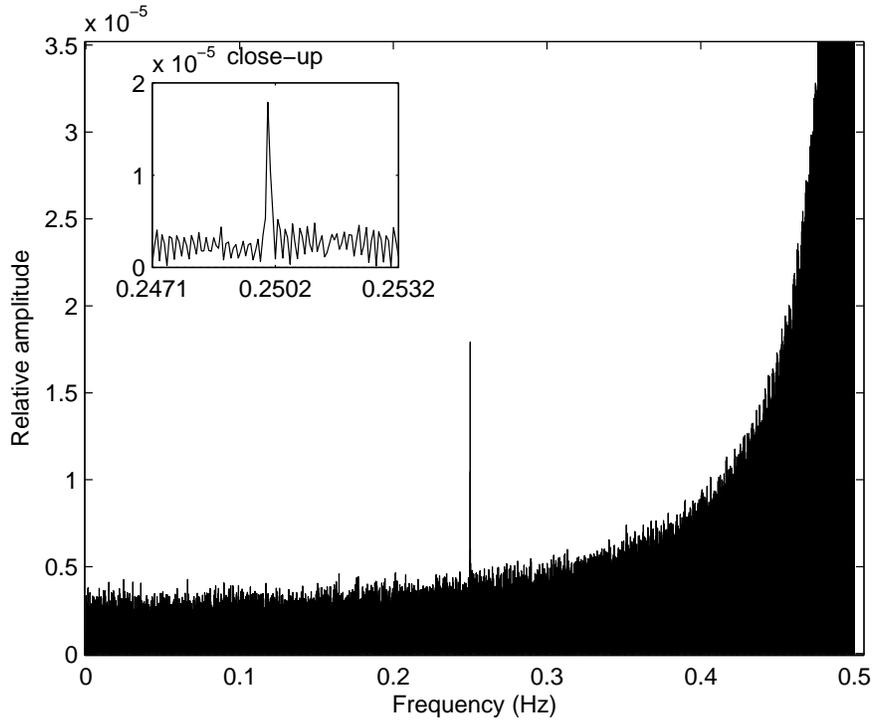}
\caption{The Fourier transform of the relative signal subtracted its average for the \texttt{W}-stage of the FCC-ee.
The waveplate rotation frequency was here chosen as 0.25Hz. At the
end of the spectrum the large component at the double frequency can
be seen, while the main signal gives a distinct peak in the center, at the chosen frequency of the quarter-wave plate. The insert shows a close-up on the peak to display its narrow width, and clarity above background. This is for a 3 hour measurement. }
\label{fig:3}       
\end{figure}

In figure \ref{fig:3} is shown the result obtained for the \texttt{W}-stage of the FCC-ee, i.e.\ operation at 80 GeV. Even for a measurement as short as 3 hours, the peak arising from the change of polarisation in the magnetic field can be clearly identified.

\begin{figure}[t]
\sidecaption[t]
\includegraphics[width=\textwidth]{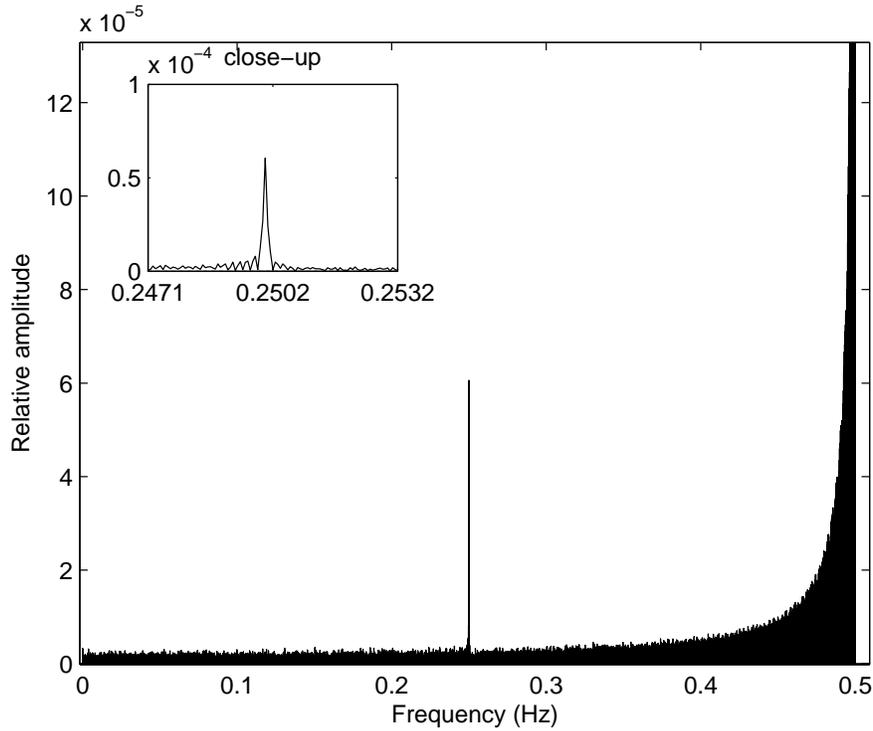}
\caption{The Fourier transform of the relative signal subtracted its average for the \texttt{tt}-stage of the FCC-ee.
The waveplate rotation frequency was here chosen as 0.25Hz. At the
end of the spectrum the large component at the double frequency can
be seen, while the main signal gives a distinct peak in the center, at the chosen frequency of the quarter-wave plate. The insert shows a close-up on the peak to display its narrow width, and clarity above background. This is for a 3 hour measurement.}
\label{fig:4}       
\end{figure}

In figure \ref{fig:4} is shown the result obtained for the \texttt{tt}-stage of the FCC-ee, i.e.\ operation at 175 GeV. Again, even for a measurement as short as 3 hours, the peak arising from the change of polarisation in the magnetic field can be clearly identified, and due mainly to the linear dependence of $\Delta$ on $\omega$, equation (\ref{eq:deltan}), an even clearer signal is obtained at the highest energy.

\section{Conclusion}
\label{sec:3}
We have shown that it is possible to measure the phenomenon of vacuum
birefringence induced by a static magnetic field with high precision,
within a quite short time frame, using Compton backscattered photons from the FCC-ee electron beam. Any outcome of such an experiment would
be interesting. Either one would measure the QED vacuum birefringence
for the first time or, in the case of an anomalous result, the experiment could point
towards new physics, for instance the existence of the axion.

\begin{acknowledgement}
UIU wishes to congratulate prof.\ W. Greiner on the occasion of his 80th birthday, and would like to thank for a very well-organized conference at the beautiful Makutsi, South Africa.
\end{acknowledgement}

\bibliography{biblio}
\bibliographystyle{unsrt}

\end{document}